\documentclass[epsfig]{article}
\usepackage{fleqn,glas,epsfig}

\usepackage{graphicx}

\def\delz{\mbox{$<\!\!\delta Z\!\!>$}}

\def\Journal#1#2#3#4{{#1} {\bf #2} (#4) #3}

\def\NPB{{\em Nucl. Phys.} B}
\def\PLB{{\em Phys. Lett.}  B}

\def\PRD{{\em Phys. Rev.} D}

\newcommand{\AmS}{{\protect\the\textfont2
  A\kern-.1667em\lower.5ex\hbox{M}\kern-.125emS}}

\begin{document}
\begin{titlepage}{GLAS-PPE/1999--09}{July 1999}

\title{Prompt photon processes in photoproduction at HERA}

\author{Sung Won Lee} 
\collaboration {On behalf of the ZEUS Collaboration}


\conference{Talk given at Photon99 Conference, Freiburg, Germany}

\note{Slides are also available from \\
http://zedy00.desy.de/\%7Eslee/report/pp99/index.html}

\begin{abstract}
We present results for the photoproduction of inclusive prompt photons and for
prompt photons accompanied by jets, measured with the ZEUS detector at HERA. 
Cross sections as a function of pseudorapidity and transverse energy
are presented for $5 < E_T^\gamma < 10$~GeV, 
$E_T^{jet} > 5$~GeV in the
centre of mass energy range 120--270 GeV.
Comparisons are made with predictions from leading
logarithm parton shower Monte Carlos and next-to-leading order QCD calculations
using currently available models of the photon structure.
NLO QCD calculations describe the shape and magnitude of the measurements
reasonably well. 
\vspace{1pc}
\end{abstract}

\end{titlepage}


\section{Introduction}

Isolated high transverse energy (``prompt'') photon processes at HERA (fig. 1) 
could yield information about the quark and gluon content of the photon,
together with the gluon structure of the proton \cite{GV}.
The particular virtue of prompt photon processes is that 
the observed final state photon emerges directly from a QCD diagram without
the subsequent hadronisation which complicates the study of high $E_T$ quarks
and gluons.

The ZEUS collaboration has recently published the first observation 
at HERA of prompt photons at high transverse momentum in photoproduction 
reactions \cite{ZPP}, based on an integrated luminosity of 
6.4 pb$^{-1}$. An NLO calculation by Gordon \cite{LG} was found to be in
agreement with the ZEUS results,
and indicates the feasibility of distinguishing between different 
models of the photon structure. 

In the present study we extend our earlier study of prompt photon
production from a data sample of 37 pb$^{-1}$.
Differential cross sections are given for the final state containing
a prompt photon, and a prompt photon accompanying jet as a function
of pseudorapidity and of transverse photon energy. 

Comparison is made with several LO and NLO (next to leading order) 
predictions, with the goal of testing different proposed
hadronic structures of the incoming photon.  

\begin{figure}
\centerline{\hspace*{10mm}
\epsfig{file=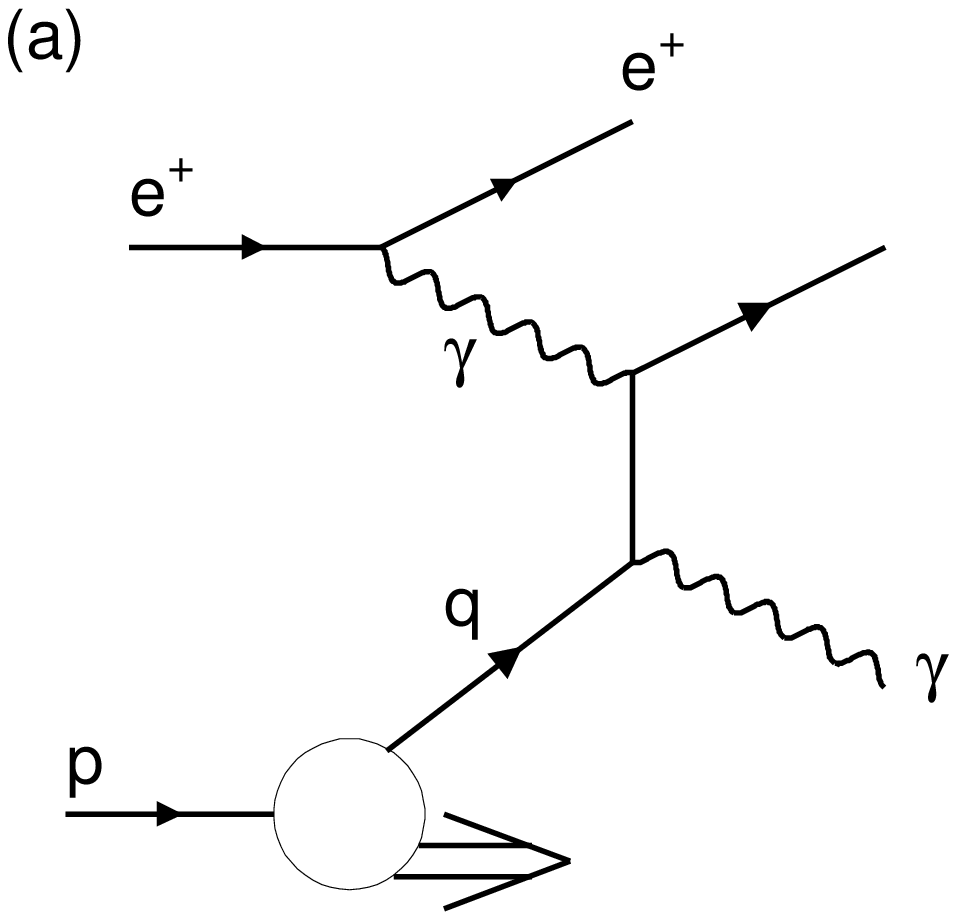,height=4.5cm,bbllx=141pt,bblly=200pt,bburx=483pt,bbury=488pt,clip=yes}
\hspace*{35mm}
\epsfig{file=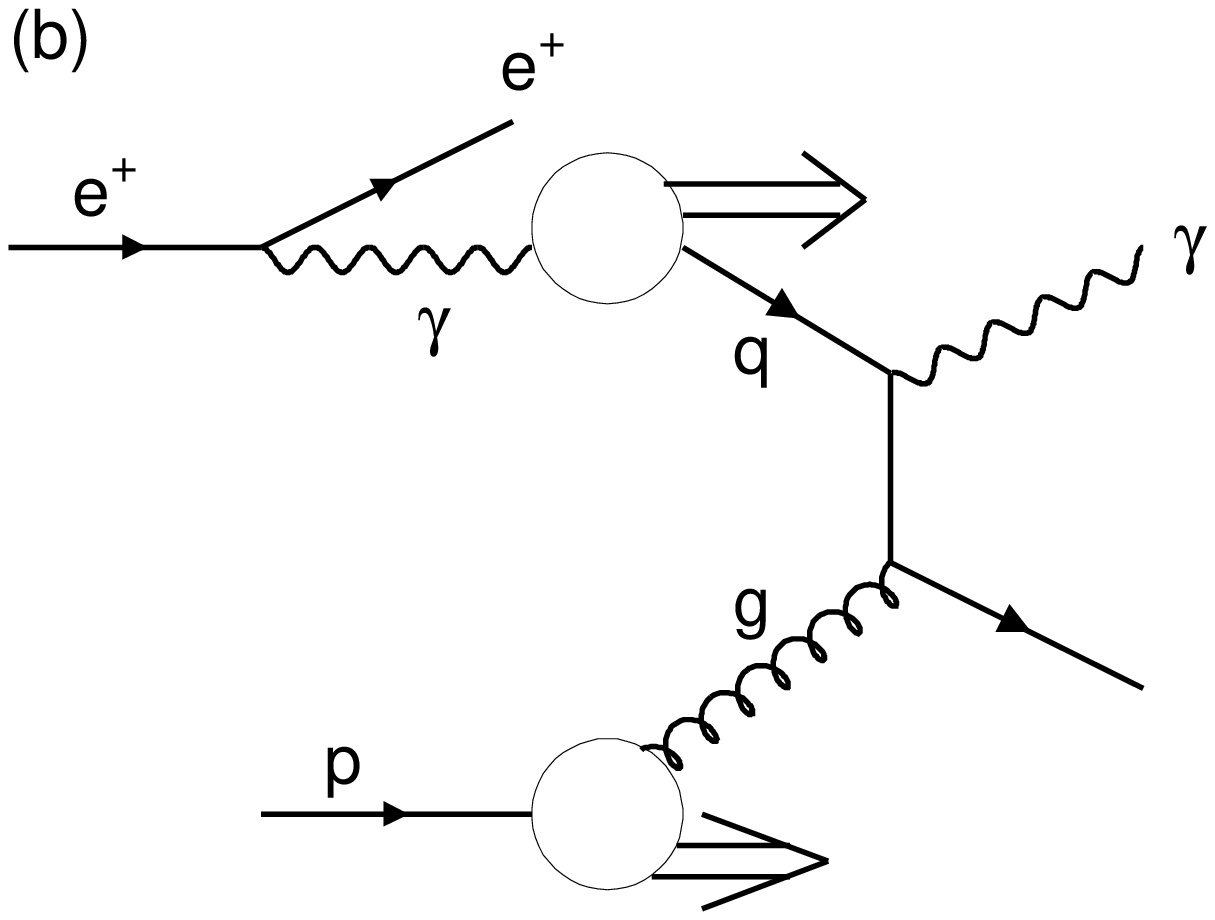,height=4.5cm,bbllx=141pt,bblly=200pt,bburx=483pt,bbury=488pt,clip=yes}
\vspace{-5mm}}
\caption{\small Example of (a) direct (pointlike) (b) resolved
(hadronic) processes in LO hard photoproduction producing an outgoing
prompt photon.}
\label{diags}\end{figure}

\section{Event Selection}

The data used here were obtained from  $e^+ p$ running in 1996--97 at HERA,
with $E_e = 27.5$ GeV, $E_p = 820$ GeV. 
The ZEUS experiment is described elsewhere~\cite{Z2}.
The major components used in the analysis are the central tracking detector
(CTD) and the uranium-scintillator calorimeter(UCAL).
Prompt photons are detected in the 
barrel section of the calorimeter (BCAL), which consists of an 
electromagnetic section (BEMC) followed by two hadronic sections; the BEMC 
consists of pointing cells of $\approx 20$ cm length and $\approx 5$ cm 
width at a minimum radius 1.23m from the beamline. 
This width is not small enough to resolve the photons from the
processes $\pi^0\to 2\gamma $, $\eta\to 2\gamma $ and $\eta\to 3\pi^0 $
on an event by event basis. It does, however, enable a partial
discrimination between single photon signals and the decay product of
neutral mesons.

A standard ZEUS electron finding algorithm was used to identify
candidate photon signals in BCAL with measured $E_T^\gamma > 4.5$ GeV.
The Energy loss in dead material to the measured photon energy
has been corrected using MC generated single photons.
This correction amounted typically to 200-300 MeV.
After the photon energy correction the events were retained for final analysis
if a photon candidates with transverse energy $E_T^\gamma > 5$ GeV was
found in the BCAL.
To identify jets,
a cone jet finding algorithm~\cite{DIJETS3} was used. 
Jet with $E_T^{jet} > 4.5$ GeV and pseudorapidity
$-1.5 < \eta^{jet} < 1.8$ were accepted with a cone radius of 1 radian,
where pseudorapidity is defined  
as $\eta = -\ln{(\tan{\theta/2})}$.
Events with an identified DIS positron were removed, 
restricting the acceptance of the present
analysis to incoming photons of virtuality $Q^2 \le1$ GeV$^2$.
The quantity $y_{JB}$, defined as the sum of $(E - p_Z)$ over all the 
UCAL cells divided by twice the positron beam energy $E_e$, provides a
measure of the fractional energy $E_{\gamma 0}/E_e$ of the interacting
quasi-real photon.
A requirement of $0.15 < y_{JB} < 0.7$ was imposed,
the lower cut removing some residual proton-gas backgrounds and the upper
cut removing remaining DIS events.
Wide-angle Compton scatters
were also excluded by this cut.

A photon candidate was rejected if a CTD track pointed within 0.3 rad of it.
An isolation cone was also imposed around photon candidates: within a cone
of unit radius in $(\eta,\,\phi)$, the total $E_T$ from
other particles was required not to exceed $0.1E_T(\gamma)$.
This greatly reduces backgrounds from dijet events with part
of one jet misidentified as a single photon ($\pi^0, \eta$, etc).
In addition, as discussed in \cite{GV}, it removes most dijet events in 
which a high $E_T$ photon radiating from a final state quark. A remainder of 
such events is included as part of the signal in the data and the theoretical
calculations.

\section{Signal/background separation}
\begin{figure}
\vspace*{-1cm}
\centerline{
\epsfig{file=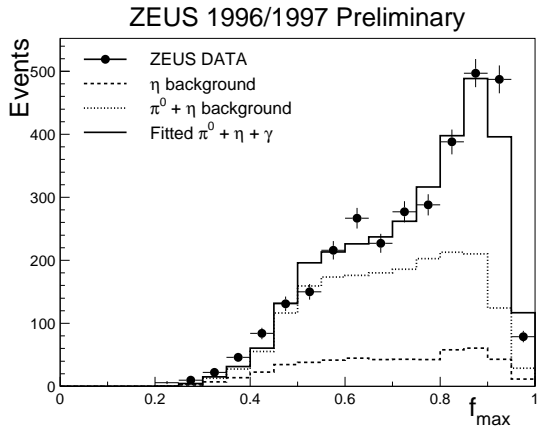,height=80mm%
}}
\vspace*{-1cm}
\caption{\small Distribution of $f_{max}$ for prompt photon candidates
in selected events, after cutting on \delz.  Also plotted are fitted
Monte Carlo curves for photons, $\pi^0$ and $\eta$ mesons with similar
selection cuts as for the observed photon candidates.}
\label{fmax}\end{figure}

A typical high-$E_T$  photon candidate in the BEMC consists  of a
cluster of 4-5 cells selected by the electron finder.
Two shape-dependent quantities were studied in order to distinguish
photon, $\pi^0$ and $\eta$ signals.  
These were (i) the
mean width $<\!\!\delta Z\!\!>$\ of the BEMC cluster in $Z$
and (ii) the fraction $f_{max}$ of the cluster energy found in the most
energetic cell in the cluster.
$<\!\!\delta Z\!\!>$\ is defined as the mean 
absolute deviation in $Z$ of the cells in the cluster, energy weighted, 
measured from the energy weighted mean $Z$ value of the cells in 
the cluster.  Its distribution shows two peaks at low $<\!\!\delta Z\!\!>$\, 
which are  identified with photons and $\pi^0$ mesons, and a tail at higher 
values. This tail quantified the
$\eta$ background; photon candidates in this region were removed.

The remaining candidates consisted of genuine high $E_T$ photons 
and $\pi^0$ and remaining $\eta$ mesons.
The numbers of candidates with $f_{max} \ge 0.75$ and $f_{max} < 0.75$
were calculated for the sample of events occurring in each bin of any
measured quantity. From these numbers, and the ratios of the
corresponding numbers for the $f_{max}$ distributions of the
single particle samples, the number of photon events in the given bin
was evaluated.  Further details of the background subtraction method
are given in~\cite{ZPP}. 
The distribution of $f_{max}$ for prompt photon candidates in
selected events is shown in fig.\ 2, well fitted to a sum of photon and
background distributions. 

\section{Results}

\begin{figure}[t]\centerline{
\epsfig{file=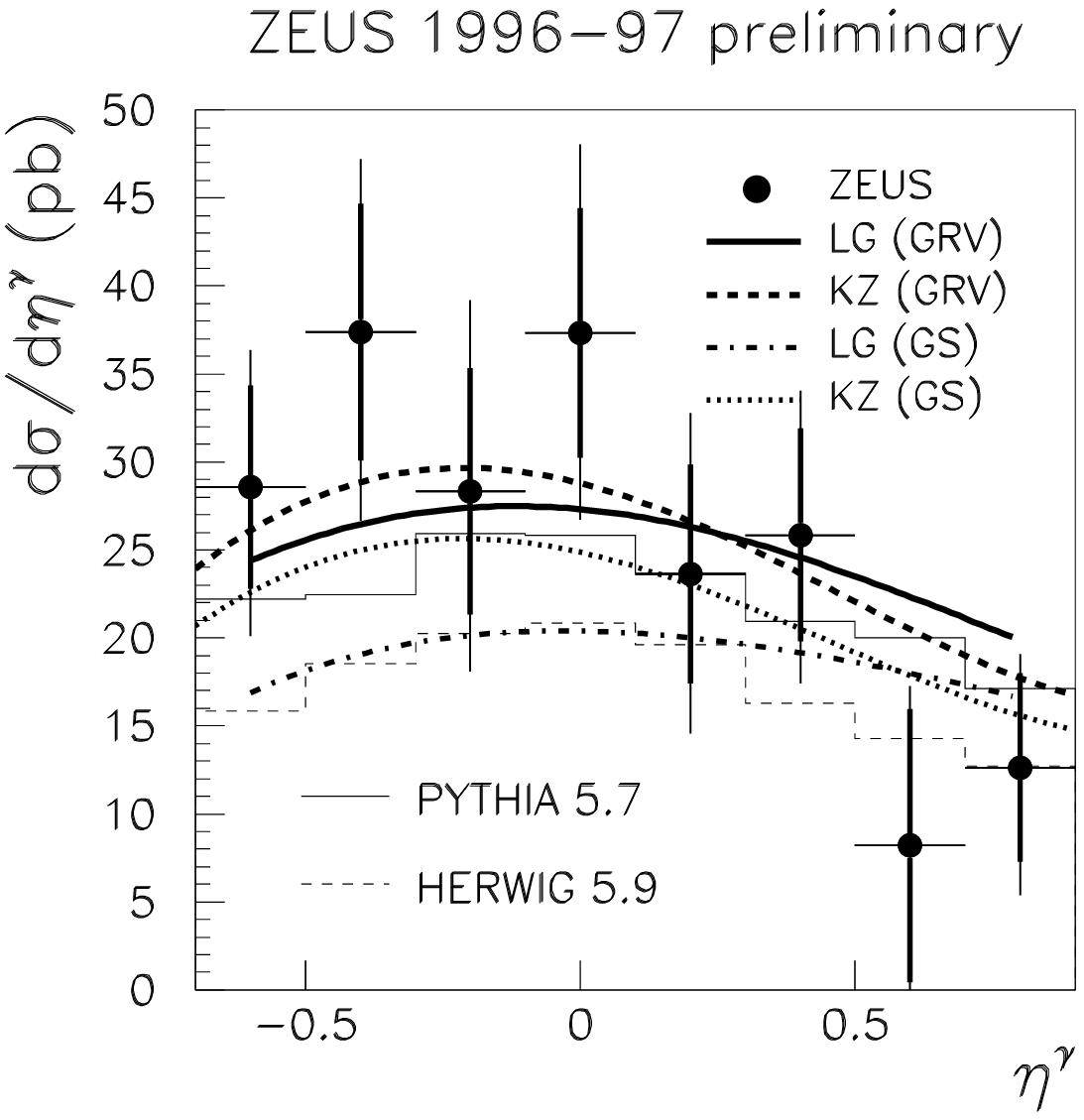,height=7cm}
\epsfig{file=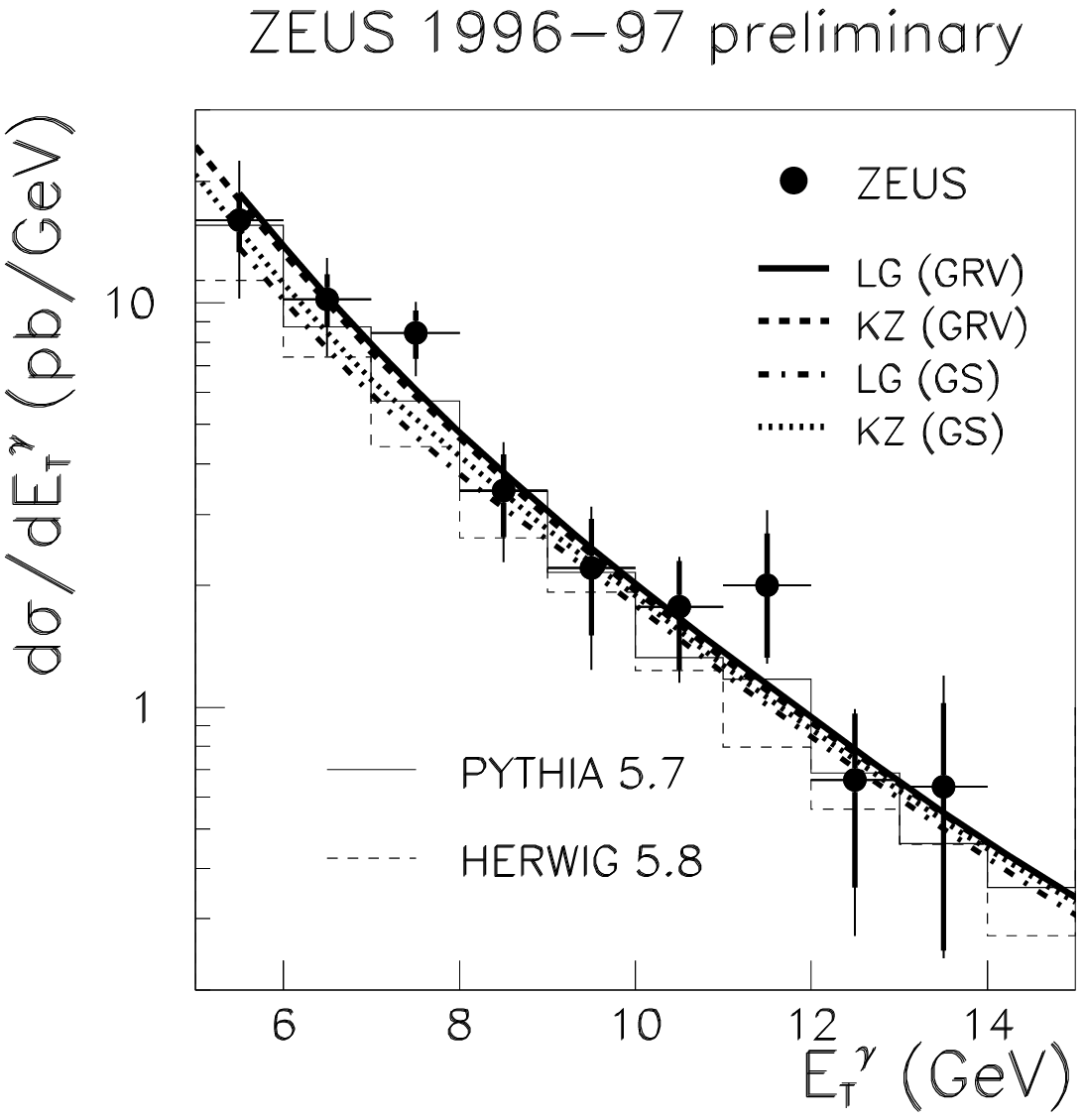,height=7cm}}
\caption{\small Differential cross sections $d\sigma/d\eta^\gamma$ for
prompt photons integrated over $5 < E_T^\gamma < 10 $ GeV,
$d\sigma/dE_T^\gamma$ for
prompt photons integrated over $-0.7 < \eta^\gamma < 0.9$.  Inner
(thick) error bars are statistical, outer include systematic added in
quadrature.  Also plotted are PYTHIA, HERWIG and NLO calculations
of LG and KZ with two different photon structures.}\end{figure}
\begin{figure}[t]
\vspace*{-0.5cm}
\centerline{
\epsfig{file=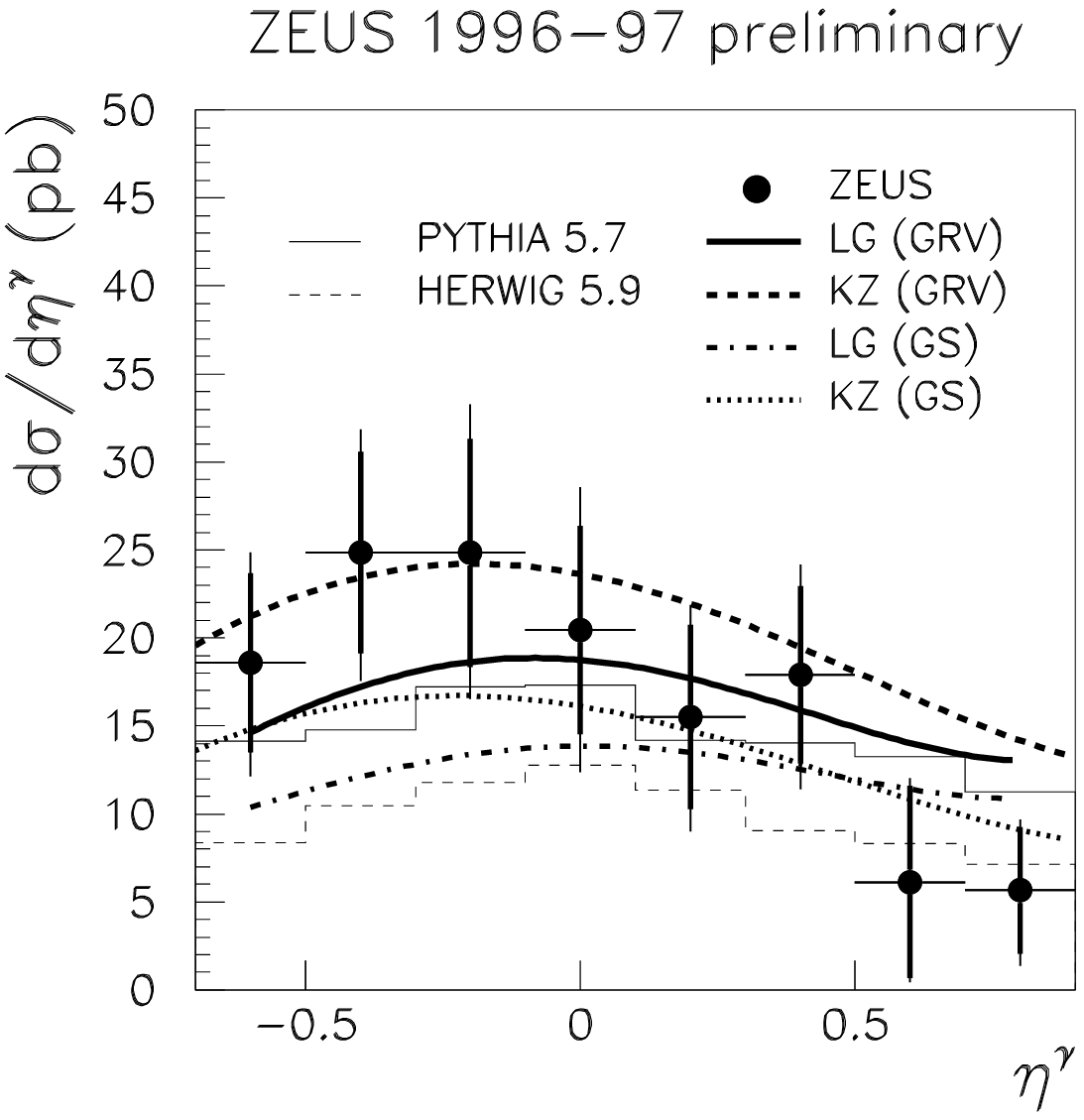,height=7cm}
\epsfig{file=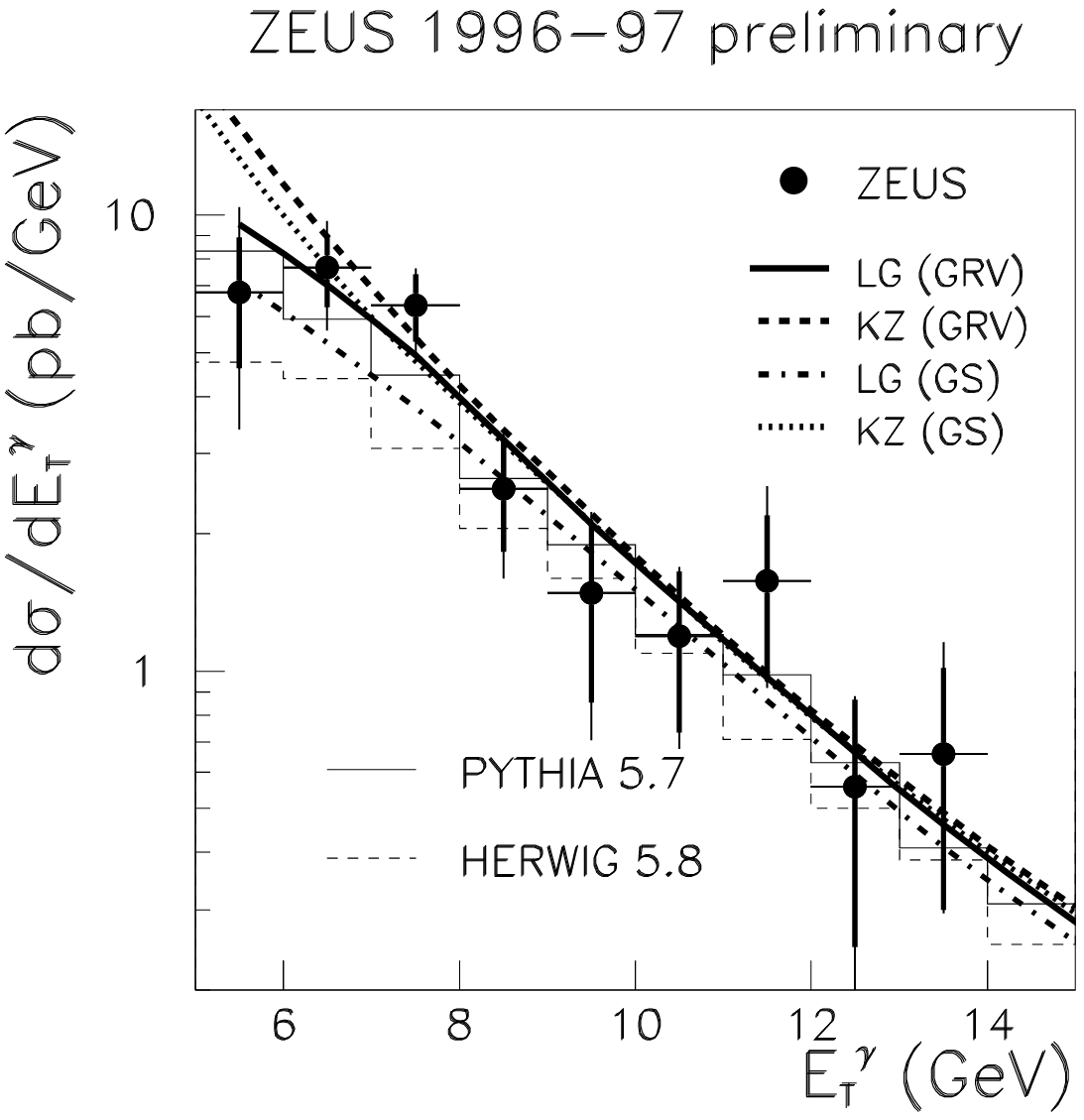,height=7cm}}
\caption{\small Differential cross sections $d\sigma/d\eta^\gamma$,
$d\sigma/dE_T^\gamma$ for prompt photons with a jet requirement.
Also plotted are PYTHIA, HERWIG and NLO calculations
of LG and KZ with two different photon structures.}
\end{figure}

We evaluate cross sections for prompt photon production corrected by
means of PYTHIA using GRV photon structures~\cite{GRV}.
A bin-by-bin factor is applied to the detector-level
measurements so as to correct to cross sections in the specified
kinematic intervals calculated in terms of the final state hadron
system photoproduced in the range $0.16 < y^{true} < 0.8$, i.e.
$\gamma p$ centre of mass energies in the range 120 -- 270 GeV.  
The virtuality of the incoming photon is restricted to the range
$Q^2 < 1$ GeV$^2$. When a jet was demanded, the hadron-level
selections $E_T^{jet} > 5$ GeV, $-1.5 < \eta^{jet} < 1.8$ were imposed.
The systematic error of 15\% were taken into account and
were finally combined in quadrature. The main contributions are
from the energy scale on the calorimeter and the background subtraction.

Fig.\ 3 shows an inclusive cross section $d\sigma/d\eta^\gamma$ 
for prompt photons in the range $5 < E_T^\gamma < 10$ GeV.
Reasonable agreement between data and MC is seen at
forward values of rapidity, but the data tend to lie above the MC
at negative rapidity. 
The data are also compared with NLO calculations
of Gordon(LG), Krawczyk and Zembrzuski(KZ)~\cite{LG,KZ} 
using the GS and GRV photon structures~\cite{GRV}.
The curves are subject to a calculational uncertainty of 5\%,
and uncertainties in the QCD scale could raise the numbers by
up to $\approx 8\%$. Away from the most forward directions, 
the LG calculation using GS tend to lie low, while the LG implementation
of the GRV photon structure give a reasonable description of the data.
KZ calculation has detailed differences from LG including a box diagram
contribution for the process $\gamma g \to \gamma g$~\cite{Combridge}.

In fig.\ 3 inclusive cross sections $d\sigma/dE_T^\gamma$ for prompt photons
in the range $-0.9 < \eta^\gamma < 0.7$ are compared to the theoretical
models. All six theoretical models describe the shape of the data well.
However the HERWIG predictions is systematically low. 
The two NLO
calculations are in better agreement with the data, 
and cannot be experimentally distinguished. Similar features
can be seen in fig.\ 4 which shows cross sections for the
production of a photon accompanied by a jet in the kinematic range
specified above.  The KZ calculation is too high at low $E_T^\gamma$, 
attributable to the lack of a true jet algorithm in this approach~\cite{KZ}.

\begin{figure}
\centerline{\hspace*{4cm}ZEUS 1996/7 PRELIMINARY\\[-11mm]}
\centerline{\hspace*{-1cm}
\epsfig{file=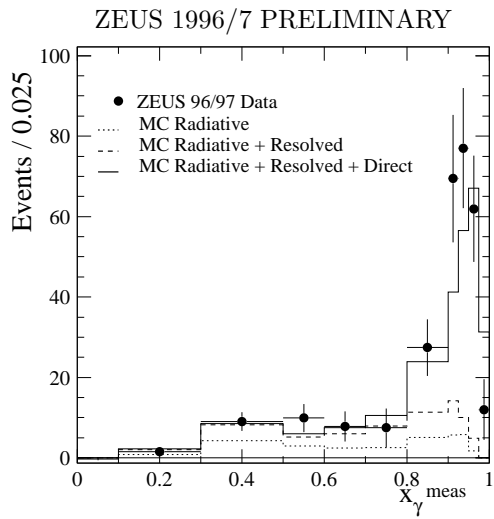,height=8cm,%
}}
\vspace*{-0.5cm}
\caption{\small $x_\gamma^{meas}$ associated with the photon+jet final state
at the detector level, compared with PYTHIA predictions.
Points = data; dotted = MC radiative; dashed = MC radiative+resolved;
solid line = MC radiative+resolved+direct. }
\end{figure}

Fig.\ 4 shows corresponding cross sections for the photon accompanied
by at least one jet. The results were corrected to hadron-level jets
in the kinematic range $E_T^{jet} > 5$GeV,
$-1.5 < \eta^{jet} < 1.8$.
In a comparison with NLO calculations from \cite{LG,KZ} 
the GS photon structure again provides a less good description of the
data overall than that of GRV.

As with the photoproduction of a dijet final state~\cite{DIJETS3}, the
information from the prompt photon and the measured jet can be used to
measure a value of $x_\gamma$, the fraction of the incoming photon
energy which participates in the hard interaction.
A ``measured" value of $x_\gamma$ at the detector level was evaluated
as $x_\gamma^{meas} = \sum(E-p_Z)/2E_e y_{JB}$, where the sum is over
the jet plus the photon.  
The  resulting distribution is 
shown in fig.\ 5 compared with PYTHIA predictions.
Reasonable agreement is seen, and
a dominant peak near unity indicates clearly the presence of the direct
process. 
Corrected to hadron level, the cross section integrated over
$x_\gamma > 0.8$ is $15.4\pm1.6(stat)\pm2.2(sys)$ pb.  
This may be compared with results from
Gordon~\cite{LG}, which vary in the range 13.2 to 16.6 pb according to
the photon structure taken and the QCD scale (approximately an 8\%
effect). 
Here, the experiment is in good agreement with the range of theoretical 
predictions but does not discriminate between the quoted models.

\section{Conclusions}

The photoproduction of inclusive prompt photons, and prompt photons 
accompanied by jets, has been measured with the ZEUS detector at HERA
using an integrated luminosity of 37~pb$^{-1}$. 
Cross sections as a function of  pseudorapidity and transverse energy
have been measured for photon transverse energies in the range
$5 < E_T^\gamma < 10$~GeV  and for jet transverse energies in the range 
$E_T^{jet} > 5$~GeV. The results are compared with 
parton-shower Monte Carlo simulations of prompt photon processes and 
with NLO QCD calculations incorporating the 
currently available parameterisations of the photon structure. NLO
QCD calculations describe the shape and magnitude of the measurements
reasonably well.


\section*{Acknowledgments}

We are grateful to L. E. Gordon, Maria Krawczyk and Andrzej Zembrzuski
for helpful conversations, and for providing theoretical calculations.



\end{document}